\documentclass[10pt,aps,twocolumn,showpacs,preprintnumbers,amsmath,amssymb,prl]{revtex4-1}
\usepackage{graphicx}% Include figure files
\usepackage{dcolumn}% Align table columns on decimal point
\usepackage{bm}% bold math
\usepackage{amsmath}
\usepackage{epstopdf}
\usepackage{color}
\usepackage{multibib}
\usepackage{comment}  % Used to suppress processing of text that we might want to bring back.

\begin{document}
%\draft
\setcitestyle{super}
\newcommand{\remark}[1]{{\color{blue}[{\bf #1}]}}
\newcommand{\red}[1]{{\color{red}#1}}
\newcommand{\blue}[1]{{\color{blue}#1}}
\newcommand{\green}[1]{{\color{green}{#1}}}

\title{Observation of nonaxisymmetric standard magnetorotational instability induced by a free-shear layer}
\author{Yin Wang$^{1}$}
\email{ywang3@pppl.gov}
\author{Fatima Ebrahimi$^{1,2}$, Hongke Lu$^3$, Jeremy Goodman$^2$, Erik P. Gilson$^1$ and Hantao Ji$^{1,2}$}
\affiliation{$^1$Princeton Plasma Physics Laboratory, Princeton University, Princeton, New Jersey 08543, USA\\
$^2$Department of Astrophysical Sciences, Princeton University, Princeton, New Jersey 08544, USA\\
$^3$Thayer School of Engineering, Dartmouth College, Hanover, New Hampshire 03755, USA}
\begin{abstract}
The standard magnetorotational instability (SMRI) is widely believed to be responsible for the observed accretion rates in astronomical disks.
It is a linear instability triggered in the differentially rotating ionized disk flow by a magnetic field component parallel to the rotation axis.
Most studies focus on axisymmetric SMRI in conventional base flows with a Keplerian profile for accretion disks or an ideal Couette profile for Taylor-Couette flows, since excitation of nonaxisymmetric SMRI in such flows requires a magnetic Reynolds number $Rm$ more than an order of magnitude larger.
Here, we report that in a magnetized Taylor-Couette flow, nonaxisymmetric SMRI can be destabilized in a free-shear layer in the base flow at $Rm\gtrsim1$, the same threshold as for axisymmetric SMRI. Global linear analysis reveals that the free-shear layer reduces the required $Rm$, possibly by introducing an extremum in the vorticity of the base flow. Nonlinear simulations validate the results from linear analysis and confirm that a novel instability recently discovered experimentally (Nat. Commun. 13, 4679 (2022)) is the nonaxisymmetric SMRI. Our finding has astronomical implications since free-shear layers are ubiquitous in celestial systems.
\end{abstract}
\date{\today}
\maketitle

The magnetorotational instability (MRI), a linear magnetohydrodynamic (MHD) instability in a differentially rotating conductive flow with a magnetic field, is thought to be the main cause of turbulence that leads to outward transport of angular momentum and inflow of mass (accretion) in astronomical disks~\cite{velikhov59,chandra60,BH91,BH98,FKR02}.
%, with magnetic winds being an alternative~\cite{Lesur21}.
An MRI-active accretion disk consists of partially ionized and magnetized plasma~\cite{EHT21} orbiting a compact massive object such as a black hole or protostar.
Such flow has a Keplerian angular velocity profile  $\Omega(r)\propto r^{-3/2}$, cylindrical radius $r$ being the distance from the central mass in the disk midplane.
%Several features of accretion disks are believed to be linked to MRI. These include turbulence~\cite{SS73,Pringle81,LL07,SH09}, the dynamo mechanism that generates and maintains an organized magnetic field component~\cite{BNST95,HGB96,ROP07,EPS09}, and perhaps large-scale oscillations~\cite{ABT06,HBFF09}.
\begin{comment}
Among MRI subvariants with different magnetic field configurations, the standard MRI (SMRI) is thought to be the most astrophysics-relevant, which requires a magnetic field parallel to the rotation axis of the flow.
This is because other versions of MRI with a pure or dominant azimuthal magnetic field, like the helical MRI~\cite{RH05,SGGRSSH06} (HMRI) and azimuthal MRI~\cite{HTR10,SGGGSGRSH14} (AMRI), are inductionless and thus incapable of generating and sustaining the needed magnetic field.
They also require rotation profiles steeper than Keplerian profile, hence are unlikely to be relevant to most astrophysical disks~\cite{LGHJ06,KSF12}.
\end{comment}
Among astrophysically relevant variants of MRI, the most straightforward is the ``standard MRI'' (SMRI)~\cite{velikhov59,chandra60,BH91}, requiring only a component of magnetic field parallel to the rotation axis and a negative radial angular velocity gradient  ($\partial\Omega/\partial r<0$).
% We call this `standard MRI'' (SMRI).
SMRI is insensitive to radial boundary conditions, which is particularly important for accretion disks as the radial boundaries are generally far away compared to the vertical scale height $L$.
% (i.e., $\parallel\vec\Omega$) , $h$.
Given finite magnetic diffusivity, $\eta$, SMRI requires a magnetic Reynolds number $Rm\equiv\Omega L^2/\eta\gtrsim 1$,
% where $L$ is an appropriate lengthscale ($L\sim h$ in disks).
and also a magnetic field not to be too strong: $V_\textsc{a}\lesssim \Omega L$,
where $V_\textsc{a}=B/\sqrt{\mu_0\rho}$ is the Alfv\'en speed based on the field strength $B$, vacuum permeability $\mu_0$ and mass density $\rho$.
These characteristics distinguish SMRI from inductionless variants of MRI that persist in the limit $Rm\to 0$ provided that dimensionless numbers scaling with $B^2/\eta$ remain nonzero (e.g., Els\"asser number $\Lambda = V_\textsc{a}^2/\eta\Omega$ or squared Hartmann number $Ha^2=V_\textsc{a}^2 L^2/\eta\nu$)\cite{RH05,SGGRSSH06,HTR10,SGGGSGRSH14}.
These inductionless variants require field strengths, field geometries, shear profiles, and/or radial boundary conditions not characteristic of most accretion disks\cite{LGHJ06,KSF12}.

Due to the limited resolution of current telescopes, SMRI cannot be confirmed by astronomical observation, but it can be accessible experimentally.
%Therefore, well-controlled laboratory experiments became the only feasible approach, such as creating it in a magnetized swirling flow of liquid metal.
In most MRI experiments, a swirling flow is created in a Taylor-Couette device consisting of two coaxial cylinders that rotate independently to viscously drive the liquid metal between them.
Ideally, such a flow has a rotation profile (ideal Couette) $\Omega(r)=a+b/r^2$, with constants $a$ and $b$ determined by the rotation speeds of the two cylinders.
According to Rayleigh’s criterion~\cite{Rayleigh17}, a purely hydrodynamic, non-MHD rotation profile with $q=-(r/\Omega)\partial\Omega/\partial r<2$ (quasi-Keplerian) is linearly stable to axisymmetric perturbations.
Experiments indicate that quasi-Keplerian hydrodynamic flow, when secondary flows due to end effects are minimized with independently rotating endcaps, is nonlinearly stable as well~\cite {JBSG06}.
With the addition of an axial magnetic field, SMRI has been demonstrated in a hydrodynamically stable liquid-metal flow~\cite{WGEGJ22}.
Previous studies of SMRI mainly focused on its axisymmetric version. This is because although nonaxisymmetric SMRI is essential for astrophysical dynamos~\cite{BNST95,HGB96,ROP07,LO08,EB16,BEB16,RL17}, it requires a higher $Rm$ in ideal-Couette flow~\cite{WGEGCWJ22,RS24}, which is out of reach of current experiments, perhaps except for the ``DRESDYN-MRI'' experiment under preparation\cite{SGGGGRSV19,Stefani24}.
On the other hand, the ideal-Couette profile cannot be fully realized in an actual device because of end effects in the presence of an axial magnetic field, and a vertical free-shear layer (Stewartson-Shercliff layer~\cite{Sherclif53,Lehnert55,Stewartson57}, SSL), is formed, creating a local maximum in $q(r)$~\cite{SRESJ12,GGJ12}.
While the stability of the SSL has been elucidated~\cite{RSGESGJ12,Roach13}, its impact on SMRI, both in the axisymmetric and nonaxisymmetric versions, has been less studied.
%Addressing this issue has particular astrophysical importance since free-shear layers are common in accretion disks, making the disk flows deviate from the Keplerian profile by creating "bumps" in the velocity or density profiles (references?).
Although constant in the idealized accretion disk discussed above, $q$ likely varies with radius in boundary layers between the disk and the central star~\cite{Pringle77,Belyaev18}, at the edges of annular gaps opened by embedded planets~\cite{Ward89,Papaloizou06}, and perhaps due to radial entropy gradient.~\cite{Klahr14}. 

Here, based on global linear analysis and nonlinear simulations of liquid-metal Taylor-Couette flows, we report for the first time that an SSL in a hydrodynamically stable axisymmetric base flow can introduce nonaxisymmetric SMRI at $Rm\gtrsim3$, a similar threshold for the axisymmetric SMRI. The nonaxisymmetric SMRI has an azimuthal mode number $m=1$ and is a standing wave in the axial direction. It leads to a global $m=1$ radial magnetic field $B_r$ in the midplane, which increases significantly with $Rm$ consistent with the recent experimental observation of an exponentially growing m=1 mode~\cite{WGEGCWJ22}.
Like $m=0$ SMRI, the SSL-induced $m=1$ SMRI also gives rise to an outward flux of axial angular momentum.
\begin{figure}
\centerline{\includegraphics[width=0.4\textwidth]{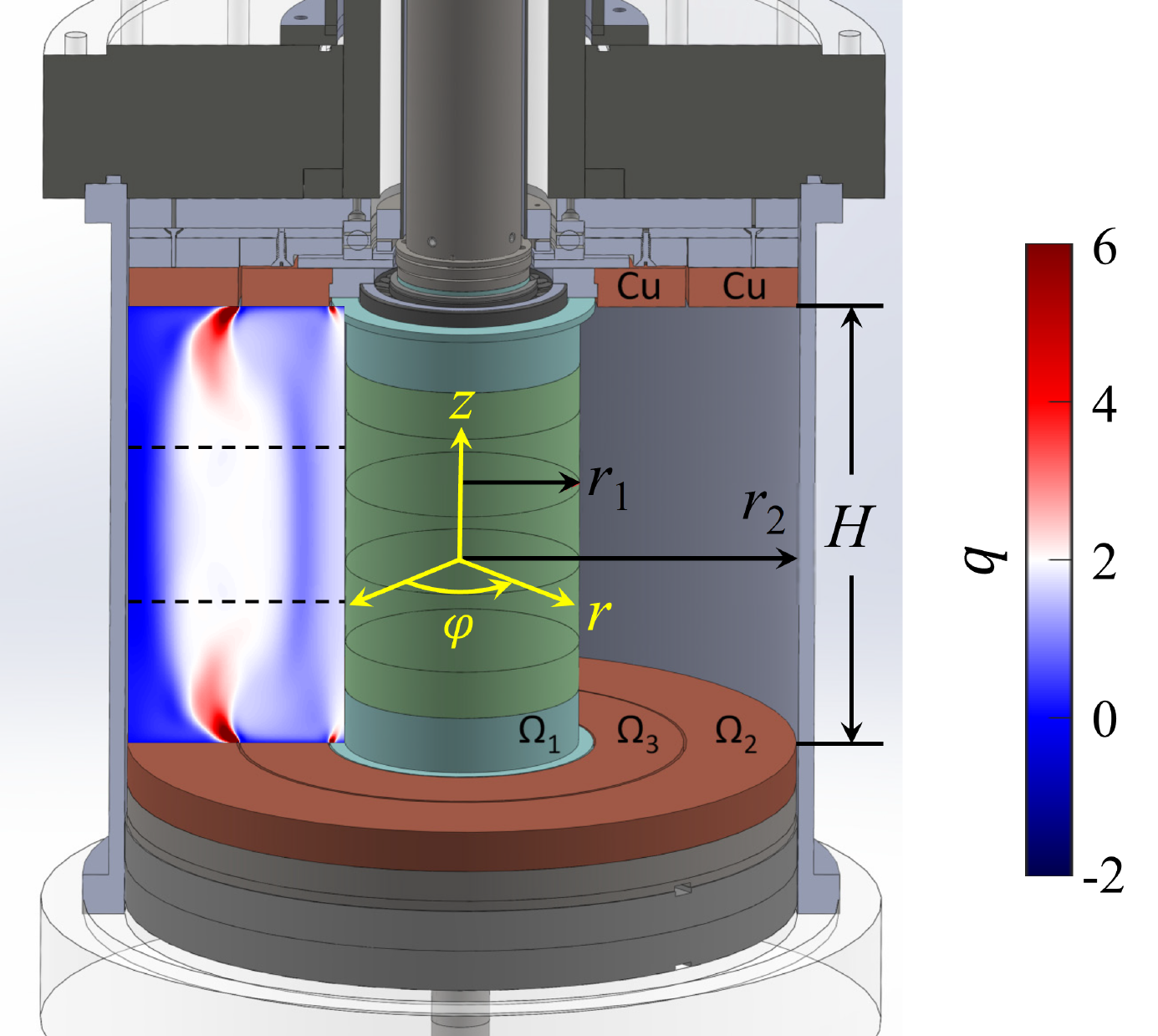}}
\caption{Sketch of the Taylor-Couette cell used in the experiment. It has three independently rotatable components: the inner cylinder ($\Omega_1$), outer-ring-bound outer cylinder ($\Omega_2$), and upper/lower inner rings ($\Omega_3$). The contour plot shows the $\varphi$-averaged shear profile $q$, in the statistically steady MHD state at $Rm=4$ and $B_0=0.2$ from 3D simulation. The cylindrical coordinate system used is shown in yellow. The $\Omega(r)$ averaged between the two horizontal dashed lines is the base flow for calculating the growth rate of the $m=1$ SMRI in Fig.~\ref {fig2}.}
\label{fig1}
\end{figure}

Details of the experimental setup have been described elsewhere~\cite{WGEGCWJ22}, and here we only mention key points (see Fig.~\ref{fig1}). 
The fluid-facing surfaces of the inner and outer cylinders have radii $r_1=7.06$ cm and $r_2=20.3$~cm, and height $H=28.0$~cm.
% The device's aspect ratio is thus $\Gamma=(r_2-r_1)/H=0.47$.
The inner cylinder is composed of five insulating Delrin rings (green).
The outer cylinder is made of stainless steel [gray; conductivity $\sigma_\mathrm{s}=1.45\times10^6$~$(\Omega\cdot m)^{-1}$].
The upper and lower end caps are copper [$\sigma_\mathrm{Cu}=6.0\times10^7$~$(\Omega\cdot m)^{-1}$] divided into two rings at $r_3=13.5$~cm,
apart from a 1-cm stainless-steel flange attached to the inner cylinder that helps further reduce the Ekman circulation~\cite{EJ14}.
The working fluid is a GaInSn eutectic alloy (Galinstan) [67\% Ga, 20.5\% In, 12.5\% Sn, density $\rho=6.36\times10^3$~kg/m$^3$, conductivity $\sigma_\mathrm{G}=3.1\times10^6$~$(\Omega\cdot m)^{-1}$], liquid at room temperature.
In the experiment, the rotation speeds of the inner cylinder ($\Omega_1$), upper/lower inner ring ($\Omega_3$), upper/lower ring ($\Omega_2$) outer cylinder ($\Omega_2$) have a fixed ratio $\Omega_1:\Omega_2:\Omega_3=1:0.19:0.58$, which has proven to minimize the hydrodynamic Ekman circulation~\cite{WGEGJ22}.
A uniform axial magnetic field $B_i \leq 4800$~G through the rotating liquid metal is provided by six coils.
Hall probes on the inner cylinder measure the local radial magnetic field $B_r(t)$ in the midplane ($z/H=0.5$) at various azimuths.
Dimensionless measures of the rotation and field strength are the magnetic Reynolds number $Rm=r_1^2\Omega_1/\eta$ and the Lehnert number $B_0=B_i/r_1\Omega_1\sqrt{\mu_0\rho}$, which are varied in the ranges $0.5\lesssim Rm \lesssim4.5$ and $0.05\lesssim B_0\lesssim1.2$.
Here $\nu$ and $\eta$ are the kinematic viscosity and magnetic diffusivity of Galinstan, and the magnetic Prandtl number $P_m=\nu/\eta=1.2\times10^{-6}$.
For each run, the device first spins up for 2 minutes. $B_i$ is then imposed, and the flow relaxes to a MHD state that is statistically steady within 2 seconds.
% The device spins up for 2 minutes, which is several times the Ekman spin-up time (approximately 40 seconds)~\cite{KJGCS04}, thus ensuring a statistically steady flow before the introduction of $B_i$.
% When $B_i$ is applied, the flow relaxes a new MHD state that is statistically steady within 2 seconds.

\begin{figure}
\centerline{\includegraphics[width=0.48\textwidth]{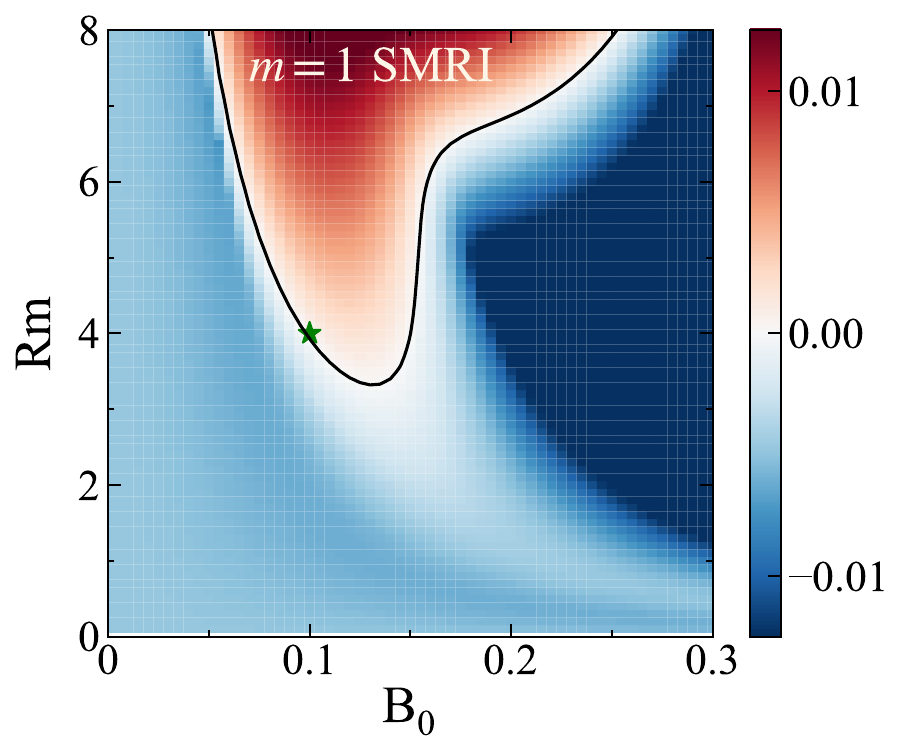}}
\caption{Criterion for nonaxisymmetric SMRI: normalized growth rate $\omega_i/\Omega_1$ of the $(m=1,n=\pm2)$ modes calculated by linear analysis in the $Rm-B_0$ plane. The black curve encloses the unstable region ($m=1$ SMRI). Insulating boundary conditions are used. The green star indicates the case of eigenfunction comparisons between linear analysis and 2D+1 simulation shown in Fig.~\ref{fig3}.}
\label{fig2}
\end{figure}

Our simulations use the open-source code SFEMaNS to solve the Maxwell and Navier-Stokes equations of an incompressible flow using spectral and finite-element methods in a fluid-solid-vacuum domain similar to our experiments. Details of the code and mesh arrangement can be found elsewhere.~\cite{GLLN09,WGEGCWJ22,WGEGJ22}
The main limitation of our simulations is the Reynolds number used, $Re=r_1^2\Omega_1/\nu\sim10^3$, versus $Re\sim10^6$ in the experiment.
Two kinds of simulations were conducted: ``3D'' nonlinear simulations containing $m=0-31$ modes and run to the saturated MHD state for comparison with experimental results in a wide range of $Rm$ and $B_0$; and ``2D+1'' linear simulations that were first run to axisymmetric MHD saturation with an SSL ($m=0$ only, $B_i>0$), after which $m=1$ terms were turned on to simulate the exponential growth phase in realistic (apart from $Re$) vertical boundary conditions.
For comparison, we performed ``1D'' global linear analyses uses the open-source code Dedalus,~\cite{Dedalus}.
Since Dedalus allows only one non-periodic dimension, these calculations assume eigenmodes $\propto\exp[i(-\omega t+m\varphi+n\pi z/H)]$: i.e., periodic in $z$ with a wavelength $\lambda_z=2H/n$, $n$ being an integer, and a base flow that is purely azimuthal with angular velocity depending on radius only, $\Omega(r)$.
%which provide the growth rate and eigenfunctions of a traveling-wave eigenmode, $e^{\omega t+i(m\varphi+n\pi z/H)}$, with a mode number combination $(m,n)$ in a given base flow with angular velocity profile $\Omega(r)$. Here, the even number $n$ is related to the mode's axial wavelength $\lambda_z$ through $\lambda_z=2H/n$, and its sign indicates the axial traveling direction of the mode. 
Here $\omega=\omega_r+i\omega_i$ where $\omega_r$ and $\omega_i$ are real angular frequency and growth rate of the eigenmode, respectively. Other details of the linear analysis are given in supplementary materials (SM).

\begin{figure}
\centerline{\includegraphics[width=0.45\textwidth]{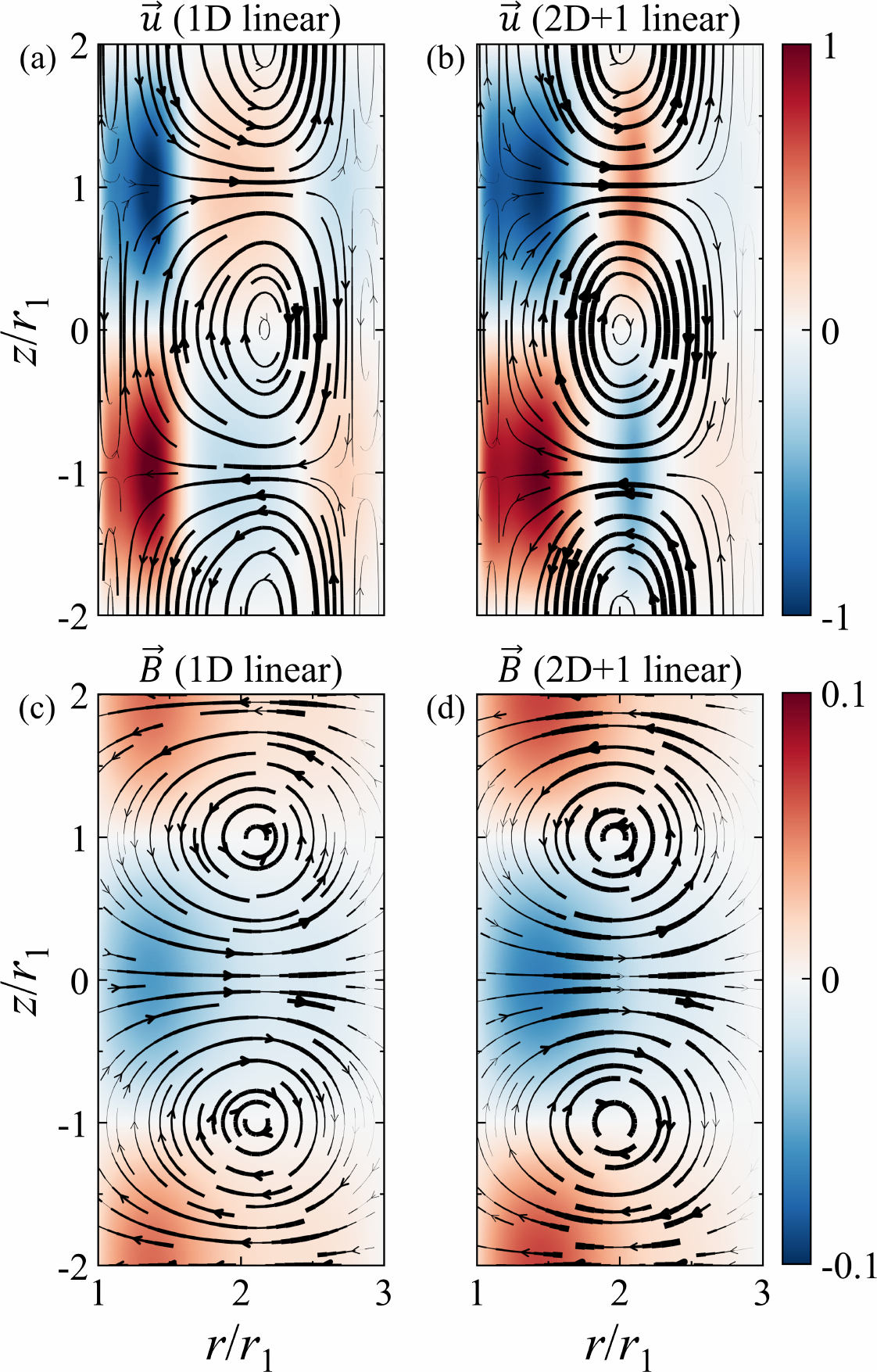}}
\caption{Global m=1 SMRI structures at $Rm=4$ and $B_0=0.1$: Poloidal cross-sectional views of eigenfunctions for $(m=1,n=\pm2)$ velocity field (a,b) and magnetic field (c,d) from 1D linear analysis (a,c) and 2D+1 linear simulation with realistic boundary conditions (b,d). Curves with arrows represent the poloidal streamlines/fieldlines constructed from the radial and axial components, and the linewidth is proportional to the local strength of the in-plane velocity/magnetic field. Color plots show the azimuthal component.}
\label{fig3}
\end{figure}

% Our 3D simulation uses the code SFEMaNS, which solves the coupled Maxwell and Navier-Stokes equations using spectral and finite-element methods in a fluid-solid-vacuum domain modeled on our experiment~\cite{GLLN09}.
% Its main difference from the experiment is that the Reynolds number $Re=r_1^2\Omega_1/\nu=10^3$, versus $Re\sim10^6$ in the experiment.
% The simulation also has two stages: it is first run without a magnetic field to a statistically steady hydrodynamic state, followed by the imposition of $B_i$ and continuing until the MHD state saturates.
% Plots in Fig.~\ref{fig1} show that except in small regions adjacent to the endcap ring gaps, $q<2$, indicating hydrodynamically stable (quasi-Keplerian) flow, as desired. Other details of numerical simulation are given in Appendix.

\begin{figure}
\centerline{\includegraphics[width=0.48\textwidth]{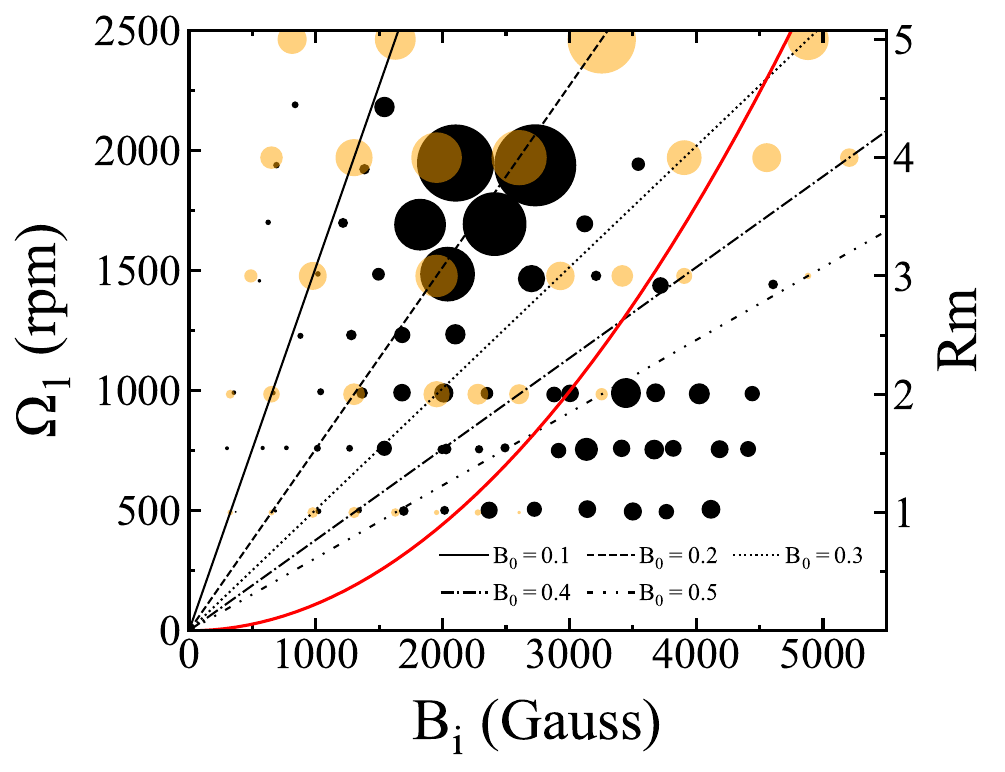}}
\caption{Bubble plot of the amplitude of the saturated nonaxisymmetric $B_r$ from experiments (black) and 3D simulations (orange).
%in the $\Omega_1$-$B_i$ plane with $Rm$ shown on the right.
The bubble size is proportional to the amplitude. Straight lines show constant Lehnert number $B_0$. The red curve represents Els\"asser number $\Lambda=1$. The experimental data were adopted from Ref.~\cite{WGEGCWJ22}. The simulation data are the volume average of the $(m=1,n=\pm2)$ modes in $B_r$.}
\label{fig4}
\end{figure}

As depicted in the color plot in Fig.~\ref{fig1}, an SSL forms at the joints of the end-cap rings after the imposition of the magnetic field, appearing as a local maximum in $q(r)$.
This corresponds to a local minimum in the flow's vorticity, $\xi(r)=(2-q)\Omega$.
In our linear analysis, the base flow $\Omega(r)$ is averaged between the two dashed lines to avoid hydrodynamic instabilities (Fig.~S1 in SM).
As shown in Fig.~\ref{fig2}, in such a hydrodynamically stable base flow, the $(m=1,n=\pm2)$ modes become unstable for $Rm\gtrsim3.5$ and $0.05\lesssim B_0\lesssim0.3$, confirming that it is the $m=1$ SMRI.
Also, near the onset, this unstable $m=1$ SMRI has a frequency of the rotation frequency of the SSL, implying that $m=1$ SMRI originates from the latter.
Notably, although the mode remains stable ($\omega_i<0$) at $Rm<3.5$, a moderate magnetic field leads to an enhanced growth rate that persists in the small-$Rm$ limit. This indicates the presence of an inductionless~\cite{JG23} branch of the SSL-induced $m=1$ SMRI, which scales with the Hartmann number $Ha$ instead of $B_0$ (Fig.~S2b in SM). Details will be reported in a future publication.
%Further investigation into it is on the way.}
For the same base flow, the $(0,2)$ mode ($m=0$ SMRI) requires $Rm\gtrsim3$ and $0.05\lesssim B_0\lesssim0.3$, a parameter space very close to the $m=1$ SMRI (Fig.~S2a in SM).
The results do not differ significantly for insulating and conducting cylinders.
In contrast, no local extrema exist in the $q(r)$ or $\xi(r)$ of an ideal Couette profile having the same $\Omega_2/\Omega_1$ (Fig.~S1 in SM), and the corresponding $m=0$ and $m=1$ SMRI requires $Rm\gtrsim10$ and $Rm\gtrsim25$, respectively, as shown in Fig.~S3 in SM.
Compared to the ideal Couette profile, the reduction in the minimum $Rm$ required for $m=0$ SMRI is due to the higher $q$ (ratio of shear to rotation) in the SSL~\cite{JGK01}. On the other hand, it is most likely that the reduction in $Rm$ for $m=1$ SMRI is caused by the local minimum in the base flow's vorticity, since no substantial reduction of $Rm$ is found in an ideal Couette base flow having a larger $q$.
The linear analysis assumes periodicity in $z$, unlike the simulations and the experiment itself.
Nonetheless, as shown by Fig.~\ref{fig3}, the superposition of the marginally unstable $(m=1,n\pm2)$ eigenfunctions from 1D linear analysis (green star in Fig.~\ref{fig2}) agrees remarkably well with those from the 2D+1 simulation using a two-dimensional base flow and physical end caps. %$m=1$ and $\lambda_z=H$ mode (the sum of $(1,2)$ and $(1-2)$) in its linear stage from the 2-mode simulation, which has a standing wave structure that gives rise to an anti-node of $B_r$ at the midplane
This confirms that the $m=1$ SMRI exists in an actual device, in which the axial boundaries impose a reflection symmetry about the midplane that leads to its standing wave structure. 
3D simulations further reveal that this structure persists in the saturated MHD state and becomes dominant over other nonaxisymmetric modes.
% the superposition of two oppositely traveling waves having the same amplitude that corresponds to the $(1,2)$ and $(1,-2)$ modes in the linear calculation.

\begin{figure}
\centerline{\includegraphics[width=0.45\textwidth]{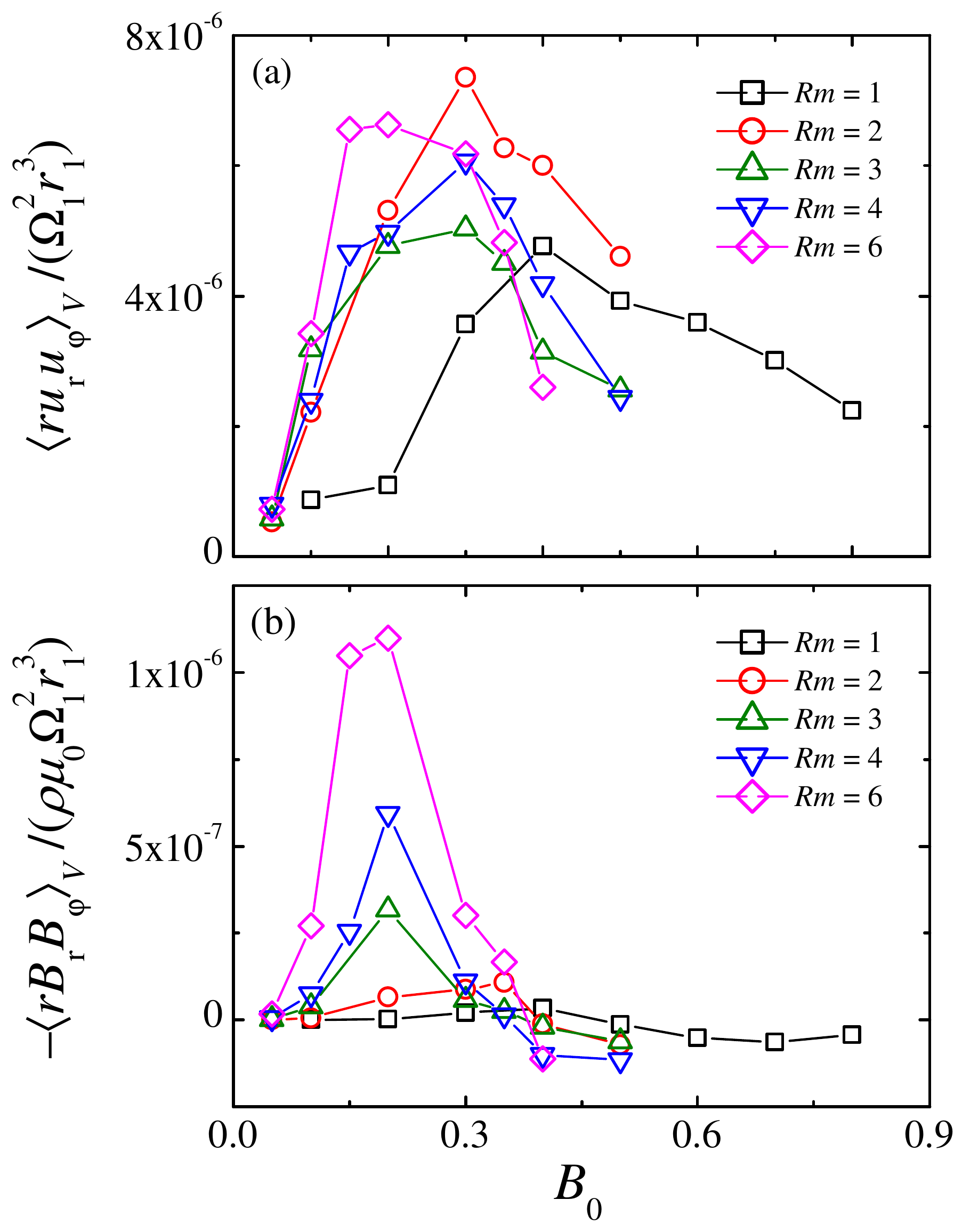}}
\caption{(a) Radial angular momentum flux $\langle ru_ru_\varphi\rangle/(r_1^3\Omega_1^2)$ contributed by the $(m=1,n=\pm2)$ mode in the velocity field, as a function of $B_0$ at various values of $B_0$ from 3D simulations.
(b) Corresponding  radial angular momentum flux $-\langle rB_rB_\varphi\rangle/(\rho\mu_0r_1^3\Omega_1^2)$ contributed by the magnetic field.
The data in both panels are time and volume averages in the saturated MHD state.}
\label{fig5}
\end{figure}

In the experiment, after the imposition of the axial magnetic field $B_i$, a linear nonaxisymmetric MHD instability with a dominant $m=1$ structure has recently been observed in the experimentally measured $B_r$ at the midplane~\cite{WGEGCWJ22}. As shown in Fig.~\ref{fig4}, this instability's amplitude has similar $Rm$ and $B_0$ dependencies as the $(m=1,n=\pm2)$ modes from 3D simulations: both become prominent at $Rm\gtrsim 3$ and $0.1\lesssim B_0\lesssim0.3$. 
In our system, the SSL becomes hydrodynamically unstable with $n=0$ once the Els\"asser number $\Lambda=B_i^2/\mu_0\rho\eta(\Omega_3-\Omega_2)>1$ (red curve)\cite{RSGESGJ12}.
The prominent bubbles have $\Lambda<1$, suggesting they are not SSL instability~\cite{WGEGCWJ22}.
As such, the consistency between the experiment and the 3D simulation confirms that the observed MHD instability is the nonaxisymmetric SMRI, as predicted by the linear analysis.

The $m=1$ SMRI induces a radial angular momentum flux through correlations between the radial and azimuthal components of velocity or magnetic fields---i.e., Reynolds or Maxwell stress.  
A normalized form of the total flux ~\cite{BH98} is:
\begin{equation}
F_r = \frac{\langle ru_ru_\varphi\rangle_V}{\Omega_1^2r_1^3}-\frac{\langle rB_rB_\varphi\rangle_V}{\rho\mu_0\Omega_1^2r_1^3},
\label{eq1}
\end{equation}
%where $\vec{u}$ and $\vec{B}$ are the velocity and magnetic field,
in which $\langle...\rangle_V$ represents averaging over the entire fluid.
The saturated $(m=1,n=\pm2)$ components of velocity and magnetic fields are used to calculate $F_r$ in Eq.~\ref{eq1}.
Figure~\ref{fig5}a shows that, for all $Rm$ studied, the Reynolds stress $\langle ru_ru_\varphi\rangle_V/\Omega_1^2r_1^3$ is always positive, first increasing and then decreasing with an increase of $B_0$. This reveals that the $m=1$ SMRI prompts an outward angular momentum flux in the velocity field, as for the axisymmetric SMRI. 
Unlike the amplitude of $B_r$
%that is enhanced by $m=1$ SMRI at $Rm\gtrsim3$ and $0.1\lesssim B_0\lesssim0.3$ 
(see Fig.~\ref{fig5}~(a)), the Reynolds stress does not increase significantly with increasing $Rm$ beyond $Rm\approx 2$, perhaps due to the fixed $Re$ studied here. 
However, the enhancement of the Reynolds stress (compared to $Rm\lesssim 1)$) occurs over a broader range than that of $B_r$; this might be caused by residual secondary flows such as Ekman-Hartmann circulation~\cite{SR07,WJGEGJL16}.
Figure~\ref{fig5}b shows the corresponding Maxwell term. Similarly, the $(1,\pm2)$ modes in the magnetic field give rise to an outward angular momentum flux in the $m=1$ SMRI unstable regime at $Rm\gtrsim3$ and $0.1\lesssim B_0\lesssim0.3$. 
Unlike the Reynolds stress, the Maxwell stress varies with $Rm$ and $B_0$ in much the same way as $B_r$ amplitude (Fig.~\ref{fig5}), and is more consistent with the growth rate shown in Fig.~\ref{fig2}.
This suggests that the magnetic field is a better diagnostic than the velocities because SMRI is an MHD instability, whereas the velocity can be confounded by hydrodynamic effects.

It has been reported that $m=1$ SMRI is observed in spherical Taylor-Couette experiments using liquid sodium, where the outer sphere is stationary and the inner sphere is rotating~\cite{SMTHDHAL04}. Although a free shear layer is present in those experiments, a stationary outer sphere leads to hydrodynamically unstable base flows that introduce other flow modes and even turbulence, complicating the measurements~\cite{GJG11}. In particular, the frequency of the free-shear layer in the spherical system is significantly higher than that of the nonaxisymmetric mode, meaning that the latter is unlikely to be excited via the nonaxisymmetric SMRI mechanism discussed here. 
%The mechanism of non-axisymmetric SMRI induced by a free shear layer is also different from a planar Couette flow, where the base flow has no local extrema and requires an infinitesimal gap between the two cylinders, which is not feasible in any practical experiment~\cite{OVBSBLB20}.

To summarize, we have presented the first confirmation of nonaxisymmetric SMRI with $m=1$ azimuthal structure in liquid-metal Taylor-Couette flow using combined theoretical, numerical, and experimental methods. Linear theory predicts that the $m=1$ SMRI can be introduced at $Rm\gtrsim3$ via a free-shear layer in a hydrodynamically stable base flow, which gives rise to a local minimum in the vorticity profile. Numerical simulations confirm the theoretical results in an actual device with vertical end caps and further reveal that the $m=1$ SMRI has a standing-wave structure in the poloidal cross-section, introducing a prominent radial magnetic field in the midplane. It also reveals that the $m=1$ SMRI introduces an outward angular momentum flux in both velocity and magnetic fields, just like the axisymmetric SMRI. By obtaining a good agreement between experiments and simulations for the $Rm$ and $B_0$ dependence of the amplitude of the nonaxisymmetric radial magnetic field, the existence of $m=1$ SMRI is thus confirmed.

Further efforts are worth making to understand the essential underlying physics of the free-shear-layer-induced SMRI by developing a unifying theoretical framework capable of explaining SMRI in different base flows. Its relationship with other local extremum-induced MHD instabilities also needs to be elucidated, such as the Rossby wave instability in the presence of magnetic fields~\cite {CTYLLY24}, or curvature effects~\cite{EP22}.
The $n=2$ nature of the nonaxisymmetric SMRI will be resolved using vertical Hall probe arrays, as well as simulations with $Re$ closer to experiments will be explored, perhaps with an entropy-viscosity method in SFEMaNS~\cite{GPP11}.

Digital data associated with this work are available from
DataSpace at Princeton University~\cite{data}.

This research was supported by U.S. DOE (Contract No. DE-AC02-09CH11466), NASA (Grant No. NNH15AB25I), NSF (Grant No. AST-2108871), and the Max-Planck-Princeton Center for Plasma Physics (MPPC). E. G. and H. J. acknowledge support by S. Prager and Princeton University. The authors thank Professor Jean-Luc Guermond and Professor Caroline Nore for their permission and assistance in using the SFEMaNS code.

\end{document}

%\draft
\setcitestyle{super}
\renewcommand{\thefigure}{S\arabic{figure}}
\newcommand{\red}[1]{{\color{red}#1}}

\title{Observation of nonaxisymmetric standard magnetorotational instability induced by a free-shear layer: Supplementary Material}
\author{Yin Wang$^{1}$}
\email{ywang3@pppl.gov}
\author{Fatima Ebrahimi$^{1,2}$, Hongke Lu$^3$, Jeremy Goodman$^2$, Erik P. Gilson$^1$ and Hantao Ji$^{1,2}$}
\affiliation{$^1$Princeton Plasma Physics Laboratory, Princeton University, Princeton, New Jersey 08543, USA\\
$^2$Department of Astrophysical Sciences, Princeton University, Princeton, New Jersey 08544, USA\\
$^3$Thayer School of Engineering, Dartmouth College, Hanover, New Hampshire 03755, USA}
\begin{abstract}

\end{abstract}
\date{\today}
\maketitle

\begin{figure}[h]
\centerline{\includegraphics[width=0.98\textwidth]{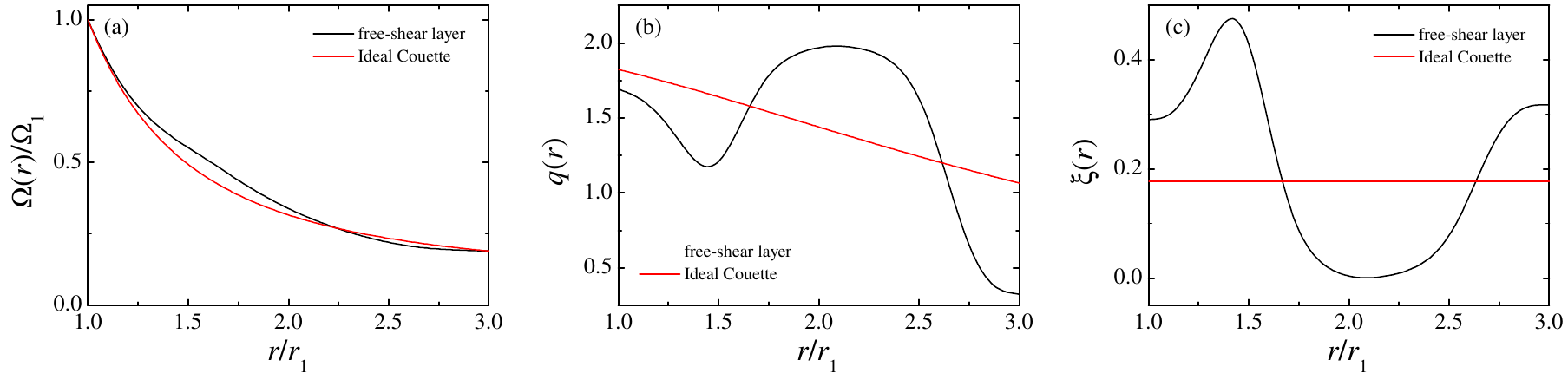}}
\caption{The radial profiles of normalized angular velocity $\Omega(r)/\Omega_1$ (a), shear rate $q(r)$ (b), and vorticity $\xi(r)$ (c) of base flows used in the linear analysis in the main text. Black curves correspond to the base flow containing a free-shear layer. Red curves correspond to an ideal Couette profile having the same $\Omega_2/\Omega_1$.}
\label{figS1}
\end{figure}

\quad

\begin{figure}[h]
\centerline{\includegraphics[width=0.98\textwidth]{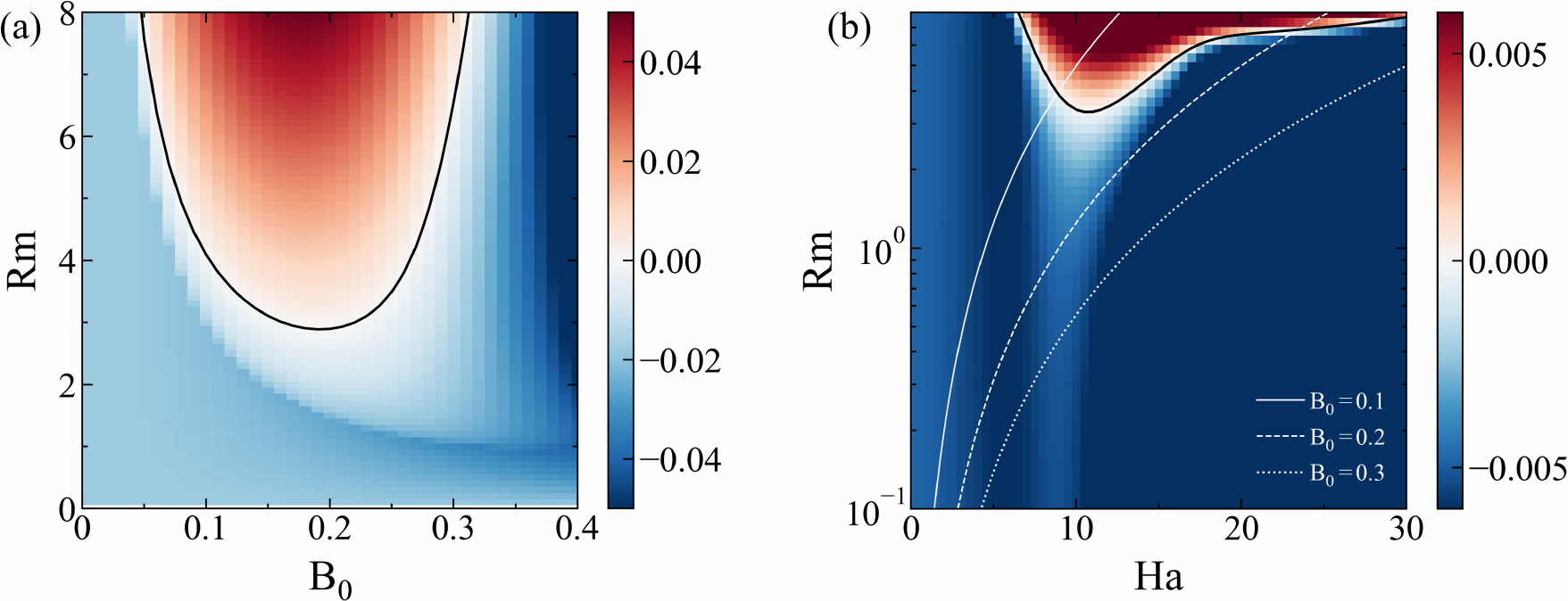}}
\caption{(a) Normalized growth rate $\omega_i/\Omega_1$ of the axisymmetric $(m,n)=(0,2)$ mode in the $Rm-B_0$ plane from global linear analysis. (b) Corresponding normalized growth rate $\omega_i/\Omega_1$ of the nonaxisymmetric mode $(m,n)=(1,2)$ in the $Rm-Ha$ plane. White curves indicate contours of constant $B_0$; the data are the same as in Fig.~2 of the main text. For both panels, the black curves enclose the unstable region ($\omega_i>0$). Insulating radial boundary conditions are used. The base flow contains a free-shear layer and its characteristics are shown as black curves in Fig.~\ref{figS1}.}
\label{figS2}
\end{figure}

\clearpage

\begin{figure}[h]
\centerline{\includegraphics[width=0.98\textwidth]{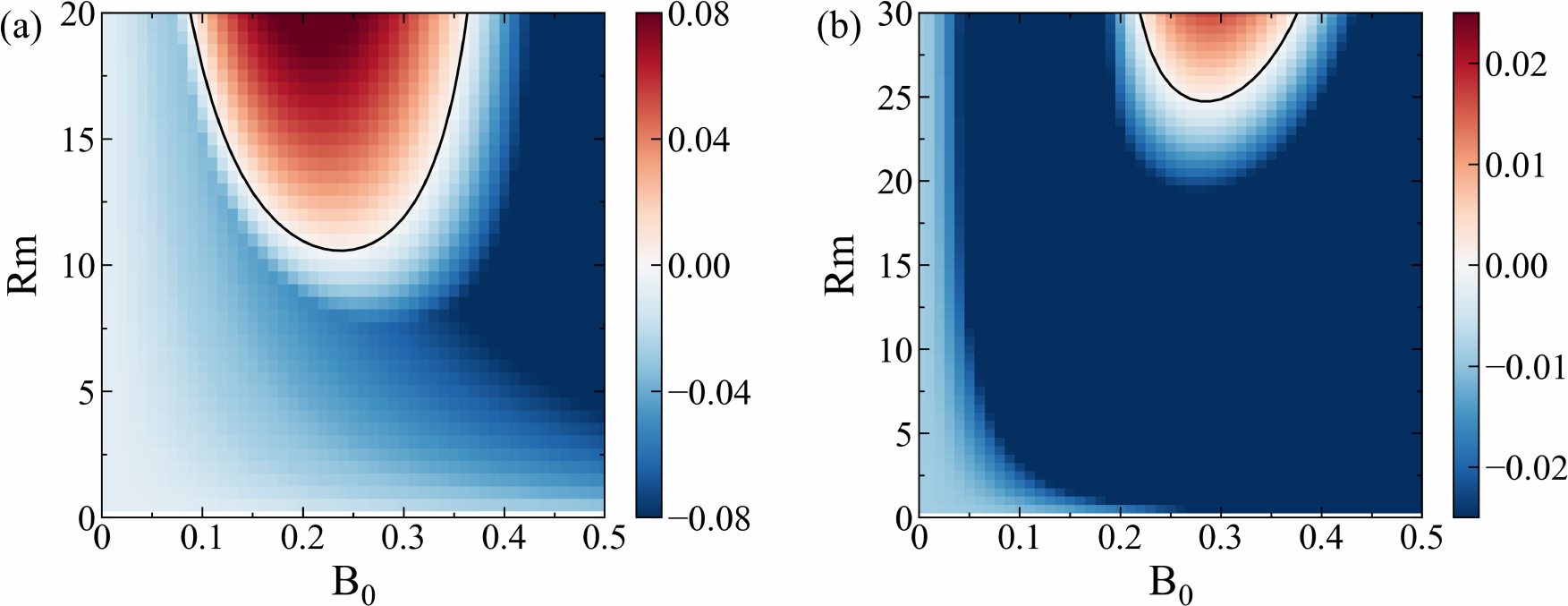}}
\caption{Normalized growth rate $\omega_i/\Omega_1$ of the axisymmetric $(m,n)=(0,2)$ mode~(a) and nonaxisymmetric mode $(m,n)=(1,2)$~(b) in the $Rm-B_0$ plane from global linear analysis. For both panels, the black curves enclose the unstable region ($\omega_i>0$). Insulating radial boundary conditions are used. The base flow has an ideal Couette profile and its characteristics are shown as red curves in Fig.~\ref{figS1}.}
\label{figS3}
\end{figure}

% \begin{figure}[h]
% \centerline{\includegraphics[width=0.56\textwidth]{figS4.pdf}}
% \caption{Normalized growth rate $\omega_i/\Omega_1$ of the axisymmetric $(0,2)$ mode from global linear analysis, in the $Rm-B_0$ plane. The black curve encloses the unstable region ($\omega_i>0$). Insulating radial boundary conditions are used. The base flow has an ideal Couette profile and its characteristics are shown as black curves in Fig.~\ref{figS1}.}
% \label{figS4}
% \end{figure}

% \begin{figure}[t]
% \centerline{\includegraphics[width=0.56\textwidth]{figS5.pdf}}
% \caption{Normalized growth rate $\omega_i/\Omega_1$ of the nonaxisymmetric $(1,2)$ mode from global linear analysis, in the $Rm-B_0$ plane. The black curve encloses the unstable region ($\omega_i>0$). Insulating radial boundary conditions are used. The base flow has an ideal Couette profile and its characteristics are shown as red curves in Fig.~\ref{figS1}.}
% \label{figS5}
% \end{figure}

\section{Global linear analysis}

% The global linear analysis deals with the linearized MHD equations for imcompressble flow
In our system, the flow dynamics is governed by the dimensionless incompressible MHD equations

\begin{equation}
    \partial_t \tilde{\mathbf{u}} + \tilde{\mathbf{u}} \cdot \nabla \tilde{\mathbf{u}} + \nabla \tilde{P} -  \tilde{\mathbf{B}} \cdot \nabla \tilde{\mathbf{B}} = \frac{1}{\mathrm{Re}} \nabla^2 \tilde{\mathbf{u}}
\label{eqS1}
\end{equation}

\begin{equation}
\partial_t\tilde{\mathbf{B}} + \tilde{\mathbf{u}} \cdot \nabla \tilde{\mathbf{B}} - \tilde{\mathbf{B}} \cdot \nabla \tilde{\mathbf{u}} = \frac{1}{\mathrm{Rm}} \nabla^2 \tilde{\mathbf{B}}
\label{eqS2}
\end{equation}

\begin{equation}
    \nabla \cdot \tilde{\mathbf{u}} = \nabla \cdot \tilde{\mathbf{B}} = 0,
\label{eqS3}
\end{equation}
where $\tilde{\mathbf{u}}$, $\tilde{\mathbf{B}}$ and $\tilde{P}$ are, respectively, the dimensionless velocity, magnetic field, and pressure. In Eqs.~\ref{eqS1}-\ref{eqS3}, the characteristic length, time, and magnetic field used for non-dimensionalization are $r_1$, $\Omega_1^{-1}$ and $\Omega_1r_1\sqrt{\rho\mu_0}$.
Eqs.~\ref{eqS1}-\ref{eqS3} are the governing equations of our nonlinear simulations~\cite{WGEGCWJ22}.
Eqs.~\ref{eqS1}-\ref{eqS3} can be linearized about a steady-state solution $(\tilde{\mathbf{u}}_0,\tilde{\mathbf{B}}_0,\tilde{P}_0)$, and the velocity, magnetic field, and pressure can be decomposed into the sum of such solution (background state) and a perturbation, namely, 
\begin{equation}
    \tilde{\mathbf{u}}(\mathbf{r},t)=\tilde{\mathbf{u}}_0(\mathbf{r})+\delta\mathbf{u}(\mathbf{r},t)
\label{eqS4}
\end{equation}

\begin{equation}
    \tilde{\mathbf{B}}(\mathbf{r},t)=\tilde{\mathbf{B}}_0(\mathbf{r})+\delta\mathbf{B}(\mathbf{r},t)
\label{eqS5}
\end{equation}

\begin{equation}
    \tilde{P}(\mathbf{r},t)=\tilde{P}_0(\mathbf{r})+\delta P(\mathbf{r},t)
\label{eqS6}
\end{equation}
For infinitely long cylinders, the background base flow $\tilde{\mathbf{u}}_0$ is purely azimuthal and the background magnetic field $\tilde{\mathbf{B}}_0$ is homogeneous and purely axial, namely,
\begin{equation}
    \tilde{\mathbf{u}}_0 = r\tilde{\Omega}(r)\mathbf{e}_\varphi
\label{eqS7}
\end{equation}

\begin{equation}
    \tilde{\mathbf{B}}_0 = B_0\mathbf{e}_z.
\label{eqS8}
\end{equation}
Here $\tilde{\Omega}(r)=\Omega(r)/\Omega_1$ is the normalized angular velocity profile of the base flow, and $\mathbf{e}_\varphi$ and $\mathbf{e}_z$ are the azimuthal and axial unit vectors in the cylindrical coordinate system. Substitute Eqs.~\ref{eqS4}-\ref{eqS8} into Eqs.~\ref{eqS1}-\ref{eqS3} and drop the quadratic perturbation terms, one gets the linearized equations for $(\delta\mathbf{u},\delta\mathbf{B},\delta P)$, which have the form~\cite{GJ02}
\begin{equation}
    \partial_t \delta{u_r} + \tilde{\Omega}\left(\partial_{\varphi}\delta{u_r}-2\delta{u_{\varphi}}\right)-B_0\partial_z\delta{B_r}+\partial_r\delta{p}-\frac{1}{\mathrm{Re}}\left[\left(\partial_r\partial_r^\dag+\frac{1}{r^2}\partial_{\varphi}^2+\partial_z^2\right)\delta{u_r}-\frac{2}{r^2}\partial_{\varphi}\delta{u_{\varphi}}\right]=0
\label{eqS9}
\end{equation}

\begin{equation}
    \partial_t \delta{u_{\varphi}} + \tilde{\Omega}\partial_{\varphi}\delta{u_{\varphi}}+\delta{u_r}\left(2\tilde{\Omega}+r\partial_r\tilde{\Omega}\right)-B_0\partial_z\delta{B_{\varphi}}+\frac{1}{r}\partial_{\varphi}\delta{p}-\frac{1}{\mathrm{Re}}\left[\left(\partial_r\partial_r^\dag+\frac{1}{r^2}\partial_{\varphi}^2+\partial_z^2\right)\delta{u_{\varphi}}+\frac{2}{r^2}\partial_{\varphi}\delta{u_{r}}\right]=0
\label{eqS10}
\end{equation}

\begin{equation}
    \partial_t \delta{B_r} + \tilde{\Omega}\partial_{\varphi}\delta{B_r}-B_0\partial_z\delta{u_r}-\frac{1}{\mathrm{Rm}}\left[\left(\partial_r\partial_r^\dag+\frac{1}{r^2}\partial_{\varphi}^2+\partial_z^2\right)\delta{B_r}-\frac{2}{r^2}\partial_{\varphi}\delta{B_{\varphi}}\right]=0
\label{eqS11}
\end{equation}

\begin{equation}
    \partial_t \delta{B_{\varphi}} + \tilde{\Omega}\partial_{\varphi}\delta{B_{\varphi}}-r\delta{B_r}\partial_r\tilde{\Omega}-B_0\partial_z\delta{u_{\varphi}}-\frac{1}{\mathrm{Rm}}\left[\left(\partial_r\partial_r^\dag+\frac{1}{r^2}\partial_{\varphi}^2+\partial_z^2\right)\delta{B_{\varphi}}+\frac{2}{r^2}\partial_{\varphi}\delta{B_{r}}\right]=0
\label{eqS12}
\end{equation}

\begin{equation}
    \frac{2}{r}\left(\tilde{\Omega}+r\partial_r\tilde{\Omega}\right)\left(\partial_\varphi\delta{u_r}-\delta{u_\varphi}\right)-2\tilde{\Omega}\partial_r\delta{u_\varphi}+\left(\partial_r^\dag\partial_r+\frac{1}{r^2}\partial_{\varphi}^2+\partial_z^2\right)\delta{P}=0
\label{eqS12}
\end{equation}

\begin{equation}
    \nabla \cdot \delta{\mathbf{u}} = \nabla \cdot \delta{\mathbf{B}} = 0.
\label{eqS13}
\end{equation}
Here $\partial_r^\dag=1/r+\partial_r$.
Assuming that the Eqs.~\ref{eqS9}-\ref{eqS13} are variable-separable, the solution with periodicity in the azimuthal and axial directions has the form
\begin{align}
\begin{split}
    \delta{u_r}=u_r(r)e^{i(-\omega t+m\varphi+k_zz)}, \quad
    &\delta{u_{\varphi}}=u_{\varphi}(r)e^{i(-\omega t+m\varphi+k_zz)}, \quad
    \delta{u_z}=u_z(r)e^{i(-\omega t+m\varphi+k_zz)}, \\
    \delta{B_r}=B_r(r)e^{i(-\omega t+m\varphi+k_zz)}, \quad
    &\delta{B_{\varphi}}=B_{\varphi}(r)e^{i(-\omega t+m\varphi+k_zz)}, \quad
    \delta{B_z}=B_z(r)e^{i(-\omega t+m\varphi+k_zz)}, \\   &\delta{P}=p(r)e^{i(-\omega t+m\varphi+k_zz)},
\label{eqS14}
\end{split}
\end{align}
where $k_z=n\pi r_1/H$ is the dimensionless axial wavenumber. Substitute Eq.~\ref{eqS14} into Eqs.~\ref{eqS9}-\ref{eqS13}, and by setting the prefactor of the exponentials to zero, one obtains a set of ordinary differential equations for the complex eigenfunctions $u_r(r)$, $u_\varphi(r)$, $u_z(r)$, $B_r(r)$, $B_\varphi(r)$, $B_z(r)$ and $p(r)$, with
\begin{equation}
i\left(-\omega + m \tilde{\Omega}\right) u_r - 2 \tilde{\Omega} u_\varphi  - i k_z B_0 B_r + p' - \frac{1}{\mathrm{Re}} \left[ u_r'' +\frac{u_r'}{r}-\left( \frac{1}{r^2} + \frac{m^2}{r^2} + k_z^2 \right)u_r - i\frac{2m}{r^2} u_\varphi \right] = 0,
\label{eqS15}
\end{equation}

\begin{equation}
i\left(-\omega + m \tilde{\Omega}\right) u_\varphi + \left( 2 \tilde{\Omega} + r \tilde{\Omega}' \right) u_r + i\frac{m}{r} p - i k_z B_0 B_\varphi - \frac{1}{\mathrm{Re}} \left[ u_\varphi'' +\frac{u_\varphi'}{r}-\left( \frac{1}{r^2} + \frac{m^2}{r^2} + k_z^2 \right)u_\varphi + i\frac{2m}{r^2} u_r \right] = 0,
\label{eqS16}
\end{equation}

\begin{equation}
i\left(-\omega + m \tilde{\Omega}\right) B_r - i k_z B_0 u_r - \frac{1}{\mathrm{Rm}} \left[ B_r'' +\frac{B_r'}{r}-\left( \frac{1}{r^2} + \frac{m^2}{r^2} + k_z^2 \right)B_r - i\frac{2m}{r^2} B_\varphi \right] = 0,
\label{eqS17}
\end{equation}

\begin{equation}
i\left(-\omega + m \tilde{\Omega}\right) B_\varphi - r\tilde{\Omega}'B_r - i k_z B_0 u_\varphi  - \frac{1}{\mathrm{Rm}} \left[ B_\varphi'' +\frac{B_\varphi'}{r}-\left( \frac{1}{r^2} + \frac{m^2}{r^2} + k_z^2 \right)B_\varphi + i\frac{2m}{r^2} B_r \right] = 0,
\label{eqS18}
\end{equation}

\begin{equation}
\frac{2}{r} \left( \tilde{\Omega} + r \tilde{\Omega}' \right) \left( im u_r - u_\varphi \right) - 2 \tilde{\Omega} u_\varphi' + p'' + \frac{p'}{r} - \left( \frac{m^2}{r^2} + k_z^2 \right)p = 0,
\label{eqS19}
\end{equation}

\begin{equation}
  u_r'+\frac{u_r}{r}+i\frac{m}{r}u_\varphi+ik_zu_z = B_r'+\frac{B_r}{r}+i\frac{m}{r}B_\varphi+ik_zB_z = 0.
\label{eqS20}
\end{equation}
No-slip boundary conditions for the velocity field are adopted for the inner and outer cylinder surfaces at $r=1$ and $r=r_2/r_1$, with
\begin{equation}
  u_r=u_\varphi=u_z = 0.
\label{eqS21}
\end{equation}
There are two kinds of boundary conditions for the magnetic field, conducting~\cite{GJ02}
\begin{equation}
  B_r = B_\varphi'+\frac{B_\varphi}{r} = 0 \quad \text{at both} \ r=1 \ \text{and} \ r=\frac{r_2}{r_1},
\label{eqS22}
\end{equation}
or insulating~\cite{RSS03}
\begin{align}
\begin{split}
    B_r + \frac{i B_z}{K_m(k_zr)} \left( \frac{m}{k_zr} K_m(k_zr) - K_{m+1}(k_zr) \right) = 0 \quad \text{at} \ r=1 \\
    B_r + \frac{i B_z}{I_m(k_zr)} \left( \frac{m}{k_zr} I_m(k_zr) + I_{m+1}(k_zr) \right) = 0 \quad \text{at} \ r=\frac{r_2}{r_1}.
\label{eqS23}
\end{split}
\end{align}
Here $I_m$ and $K_m$ are the modified Bessel functions of the first and second kind, respectively.

Equations~\ref{eqS15}-\ref{eqS19}, together with the boundary conditions Eq.~\ref{eqS21} and Eq.~\ref{eqS22} (or Eq.~\ref{eqS23}), compose an eigenvalue problem (EVP), which is numerically solved by the one-dimensional EVP solver of the open-source code Dedalus~\cite{BVOLB20}. For a given set of $m$, $n$, $B_0$, $Re$ and $Rm$, among all solutions, we only take the \emph{most unstable} one having the largest $\omega_i$. We fixed $Re=2000$ for the results presented in this manuscript, and an increase of $Re$ will slightly enlarge the unstable region. We use $N=192$ Chebyshev grid points for radial discretization, which is sufficient since its results have a relative difference of less than $10^{-6}$ compared with those using $N=1024$. Once $u_r$, $u_\varphi$, $B_r$, and $B_\varphi$ are obtained, $u_z(r)$ and $B_z(r)$ are calculated by Eq.~\ref{eqS20}.